\documentclass[aps, pra, showpacs, twocolumn, amsfonts, amsmath, amssymb, superscriptaddress, floatfix,float]{revtex4} 
\DeclareMathOperator{\Erfc}{Erfc}
\usepackage{amsmath}
\usepackage{graphicx}
\begin{document}

 \title{A Metal-Insulator transition induced by Random Dipoles}

\author{M. Larcher}
\affiliation{INO-CNR BEC Center and Dipartimento di Fisica, Universit\`a di Trento, 38123 Povo, Italy}

\author{C. Menotti} 
\affiliation{INO-CNR BEC Center and Dipartimento di Fisica, Universit\`a di Trento, 38123 Povo, Italy}

\author{B. Tanatar}
\affiliation{Department of Physics, Bilkent University, Bilkent, 06800 Ankara, Turkey}

\author{P. Vignolo}

\affiliation{Universit\'e de Nice - Sophia Antipolis, Institut non
  Lin\'eaire de Nice, CNRS, 1361 route des Lucioles, 06560 Valbonne,
  France}

\begin{abstract}
We study the localization properties of a test dipole feeling the
disordered potential induced by dipolar impurities trapped at random
positions in an optical lattice. This random potential is marked by
correlations which are a convolution of short-range and long-range
ones.  We show that when short-range correlations are dominant,
extended states can appear in the spectrum. Introducing long-range
correlations, the extended states, if any, are wiped out and
localization is restored over the whole spectrum. Moreover, long-range
correlations can either increase or decrease the localization length
at the center of the band, which indicates a richer behavior than
previously predicted.
\end{abstract}

\pacs{03.75.-b; 64.60.Cn; 71.23.An}
 
\maketitle
\section{Introduction}
Interference effects induced by random potentials deeply modify wave
diffusion up to even stop wave transport on a length
$\mathcal{L}_{loc}$, the localization length.  This phenomenon called
strong localization or Anderson localization \cite{Ande58,Ande85} can
be observed in classical waves such as acoustic waves \cite{Hu2008},
microwaves \cite{Laurent2007}, light \cite{Storzer2006}, as well as in
quantum waves, such as electronic \cite{Cluter1969} or matter waves in
real \cite{Billy2008,Roati2008,Kondov2011,Jendrzejewski2012} or
momentum space \cite{Moore1995,Chabe2008}.  Anderson theory relies
essentially on two main assumptions: (i) the potential has to be
$\delta$-correlated; (ii) the wave has to be noninteracting.  If these
conditions are fulfilled, Anderson localization occurs always in one
(1D) and two dimensions (2D), and depending on the disorder strength
and the energy in three dimensions (3D) \cite{Ande79}.  A
localization-delocalization transition can appear in low dimensions if
correlations and/or interactions are introduced in the systems. If on
the one hand noninteracting waves exist in nature, on the other hand
real uncorrelated potentials do not exist.  Correlated disordered
potentials can be roughly divided into three wide classes, accordingly
to the behavior of the two-point auto-correlation function $C(\ell)$
and of its Fourier transform $S(k)$.  The first class is marked by a
correlation function $C(\ell)$ which decays exponentially on a length
$\bar\ell$ with $\pi/k_{max}<\bar\ell\ll L$, $L$ being the length of
the system and $k_{max}$ the largest wavevector allowed by the system.
These potentials, called short-range correlated potentials, may
introduce resonance energies in the spectrum inducing delocalization
of a significant subset of the eigenstates.  This happens for instance
in the random-dimer model (RDM) and in its dual counterpart (DRDM)
\cite{Phil90,schaff10,Sedrakyan2011,Farchioni2012}, in which the sites
of a lattice are assigned energies $\varepsilon_a$ or $\varepsilon_b$
at random, with the additional constraint that sites of energy
$\varepsilon_b$ always appear in pairs (RDM) or never appear as
neighbors (DRDM).  The second class of correlated potentials is marked
by a spectral function $S(k)$ that is non-zero in a finite
$k$-range. This is, for instance, the case of the speckle
\cite{Gurevich2009,Luga09} where $C(\ell)$ is a sinc function.  For
these kinds of potentials, there exists a critical energy at which the
localization length increases abruptly mimicking the presence of a
mobility edge in finite-size systems \cite{Tessieri2002,Kuhl08}.  In
the third class $C(\ell)$ decreases algebraically as $\sim
1/\ell^{\beta}$, and both $C(\ell)$ and $S(k)$ are non-zero over the
whole real and $k$ spaces.  In practice there are no length scales
characterizing the disorder, which is scale-free. These potentials are
commonly called long-range correlated.  In this case, it has been
observed that correlations can have different effects depending on the
energy region under consideration. In particular, for discrete models,
a reduction of the localization length has been observed at the band
edges and conversely an enhancement at the band center has been
reported \cite{croy2011}.  In this context also the presence of
mobility edges has been claimed \cite{deMour98}, although these
results stirred some controversy \cite{commdeMoura,replydeMoura}.

\begin{figure}
\includegraphics[width=0.9\linewidth]{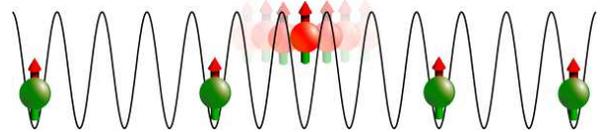}
\caption{\label{fig1} (Color online) Schematic representation of the
  physical model. Dipolar {\it impurities} (green spheres) are trapped
  at the minima of the optical lattice, occupying random
  positions. The {\it test} dipole (red upper sphere), excited to another
  internal level, feels a shallower optical potential and a disordered
  potential due to the dipolar interaction with the impurities.}
\end{figure}

Very often, especially in the case of long-range correlations, these
studies rely on toy models characterized by {\it ad-hoc} correlation
functions, creating almost no connection with possible experimental
implementations.
In this paper we propose a physical model for a random potential where
long-range and short-range correlations arise naturally from the
system itself, exploiting the properties of the dipolar interaction
\cite{Lahaye2009, Baranov2012} in ultracold atomic
\cite{Pfau2005,Laburthe2010,Lev2011,Ferlaino2012} or molecular
\cite{Ye2008,Zwierlein2012} gases. The model considers a series of
dipoles pinned at random positions at the minima of a deep optical
lattice.  The dipoles are polarized perpendicularly to the lattice
axis, so that dipole-dipole interaction is repulsive. In this way, for
low enough densities, there are no double occupancies, and, if the
dipole-dipole interaction is large enough, consecutive occupations are
avoided too.  This set of trapped dipoles, referred as {\it
  impurities}, will create a disordered potential
\cite{Vign03,Gavish2005a,schaff10} for another dipole, the {\it test
  dipole}, which is excited to a different internal level and is able
to move through the lattice (see Fig.~\ref{fig1}).  Short-range
correlations arise from the distribution of the impurities, while
long-range correlations are due to the dipolar interaction between the
test dipole and the impurities.

We study the localization properties of the test dipole in the
correlated potential realized by the impurities, highlighting the role
played by short and long-range correlations. In particular, depending
on the parameters of the model, we observe that short-range
correlations can introduce a discrete set of extended states in the
system while long-range correlations tend to restore localization and
lead to counterintuitive effects on the localization length of the
system.

The paper is organized as follows.  The model is presented in detail
in Sect.~\ref{themodel} and the Hamiltonian for the test dipole is
derived. In Sect.~\ref{thespectrum}, we study the localization
properties of the model by using a renormalization-decimation scheme
for the calculation of the localization length. Then, a detailed
discussion of the role played by short and long correlations is
presented in Sect.~\ref{sec:corr}. Finally, we draw our conclusions in
Sect.~\ref{concl}.

\section {The physical model}
\label{themodel}

We consider a very dilute gas of dipolar {\it impurities} trapped in a
deep 1D optical lattice and a single {\it test} dipole excited to a
different internal level with the same dipole moment ${\vec D}$. We
superimpose to the optical lattice a very elongated cigar-shaped
harmonic confinement, so that we can practically neglect the weak
axial harmonic confinement along the $z$ axis. We assume both for the
impurities and the test dipoles frozen radial dynamics into the lowest
state of the radial harmonic confinement

\begin{equation}
	\phi_\omega(\vec{r}_\perp)=\dfrac{1}{\sqrt{\pi}\sigma_\omega}e^{-r_\perp^2/2\sigma_\omega^2} \qquad\text{with}\qquad \sigma_\omega=\sqrt{\frac{\hbar}{m\omega}},
	\label{eq:ho_gs}
\end{equation}
where $\omega=\omega^{(I)}$ and $\omega=\omega^{(T)}$ for impurities and test particle respectively.

Furthermore, assuming that the optical lattice felt by the impurities
is so strong to completely freeze their dynamics along $z$, we can
describe the motion of the test particle of mass $m$ along the lattice
axis $z$ by the following 1D effective Hamiltonian
\begin{equation}
	H= -\dfrac{\hbar^2}{2m}\frac{\partial^2}{\partial z^2}+s E_R \sin^2(kz)+V_d(z),
	\label{eq:Hamiltonian1}
\end{equation}
where $E_R=\pi^2 \hbar^2/2m d_L^2$ is the recoil energy, $d_L$ the
lattice spacing of an optical lattice generated by a laser of
wavelength $\lambda_L=2d_L$, and $s$ the adimensional lattice depth.

The random potential $V_d(z)$ results from the dipolar interaction of
the test particle with the impurities pinned in the lattice. It is
given by the convolution of the density distribution of the impurities
$\rho(z)$ (see Eq.~\ref{eq:impurity_density}) and the effective one
dimensional dipolar interaction $U^{1D}_{dd}(z)$ (see Eq.~\ref{U1D}):

\begin{equation}
	V_d(z)=\int dz' \rho(z') U^{1D}_{dd}(z-z').
	\label{eq:dipolar_V}
\end{equation}
The density distribution $\rho(z)$ is given by the sum of the Wannier
functions $w^{(I)}(z)$ localized around the sites $\bar{l}$ occupied
by the impurities
\begin{equation}
	\rho(z)=\sum_{\bar l} |w^{(I)}(z-\bar{l}d_L)|^2.
	\label{eq:impurity_density}
\end{equation}	
The random distribution of the occupied sites $\bar{l}$ introduces disorder in the system.

The effective one dimensional dipolar potential $U^{1D}_{dd}(z)$ is
obtained after integration of the dipolar interaction $U_{dd}(\vec r)=
D^2 [1-3 ({\hat D} \cdot {\hat r})^2] /|\vec r|^3$ in the radial
directions and is given by \cite{Sinha2007}
\begin{widetext}
\begin{equation}
\label{U1D}
\begin{split}
	U_{dd}^{1D}(z)&= \int d\vec{r'}_{\perp}d\vec{r}_{\perp}|\phi_{\omega^{(I)}}{(\vec{r'}_\perp)}|^2 |\phi_{\omega^{(T)}}({\vec{r}_\perp})|^2 U_{dd}(\vec r-\vec{r'}) \\
	&=\frac{D^2}{\sigma_\perp^3} (1-3\cos^2{\alpha})
\left\{-\frac{2}{3}\delta\left(\frac{z}{\sigma_\perp}\right)+\frac{1}{2}\sqrt{\frac{\pi}{2}}e^{\frac{1}{2}\frac{z^2}{\sigma^2_\perp}}\left[\left(\frac{z^2}{\sigma^2_\perp}\right)+1\right] \Erfc\left(\frac{|z|}{\sqrt{2}\sigma_\perp}\right)-\frac{|z|}{2\sigma_\perp} \right\},
\end{split}
\end{equation}
\end{widetext}
where
$\sigma_\perp=\sqrt{(\sigma^2_{\omega^{(I)}}+\sigma^2_{\omega^{(T)}})/2}$
is the radial width of the system, $\alpha$ is the angle between the
dipole moment $\vec D$ and the $z$ axis and $\Erfc(z)$ is the
complementary error function

\begin{equation}
	\Erfc(z)=\frac{2}{\sqrt{\pi}}\int_z^\infty e^{-t^2}\,dt.
\end{equation}
Note that the final expression that we obtained for $U^{1D}_{dd}(z)$
is composed of two parts: a Dirac delta term at $z=0$ and a slowly
decaying part. One can show that at large distances $|z| \gg
\sigma_\perp$, the slowly decaying part reproduces the typical
behavior of the dipolar interaction, namely a decrease with the
inverse cubic distance $U^{1D}_{dd}(z)\sim D^2(1-3\cos^2\alpha)
/|z|^3$.
For simplicity, in the present paper, we do not include contact
interactions with the underlying idea that they can be switched off by
exploiting Feshbach resonances \cite{Koch2008}. However, they would
simply modify the strength of the $\delta$ part of the 1D potential
and provide an additional way to tune the parameters of the system.

The tight binding form of the Hamiltonian is obtained using as a basis
the set of Wannier states $w_n(z)$ for the test particle. For the case
of a single impurity pinned at site $l$ we obtain

\begin{equation}
\begin{split}
H_{l}=&\sum_n  -J\left(\mid w_n\rangle \langle w_{n+1}\mid+ \mid w_{n+1}\rangle \langle w_n\mid\right)\\
-&J^d\left(\mid w_l\rangle \langle w_{l\pm1}\mid+ \mid w_{l\pm1}\rangle 
\langle w_l\mid\right)\\
+& u^{dd}_{n-l} \mid w_n\rangle\langle w_n\mid .\\
\label{eq:tb_Hamiltonian}
\end{split}
\end{equation}
In Eq.~(\ref{eq:tb_Hamiltonian}), beyond the standard nearest
neighbour tunneling term $J$, we have included two terms due to the
dipolar interaction: the first represents a nearest neighbour dipolar
assisted hopping $J^d$, while the second contains the on-site energies
$u^{dd}_{n-l}$ at site $n$. The Hamiltonian parameters can be
calculated using the following expressions

\begin{align}
	J&=-\int w_n^*(z) \left[-\dfrac{\hbar^2}{2m}\dfrac{d^2}{dz^2}+s E_R\sin^2(kz)\right]w_{n+1}(z)\,{d}z , \nonumber  \\
	J^d&=   -\int \, w_l^*(z) w_{l+1}(z)   |w^{(I)}_l(z')|^2 U^{1D}_{dd} (z-z')\, {d}z\,{d}z' , \label{eq:tb_terms} \\
	u^{dd}_{n-l}&= \int  |w_n(z)|^2  |w^{(I)}_l(z')|^2 U^{1D}_{dd}(z-z') \, {d}z\,{d}z'.    \nonumber
\end{align}
The function $u^{dd}_{n-l}$ simply depends on the distance $|n-l|$ between the test particle and the impurity and provides the dipolar interaction between a single impurity and the test particle in the discretized formalism.
 
\begin{figure}
\begin{center}
\includegraphics[width=0.99\linewidth]{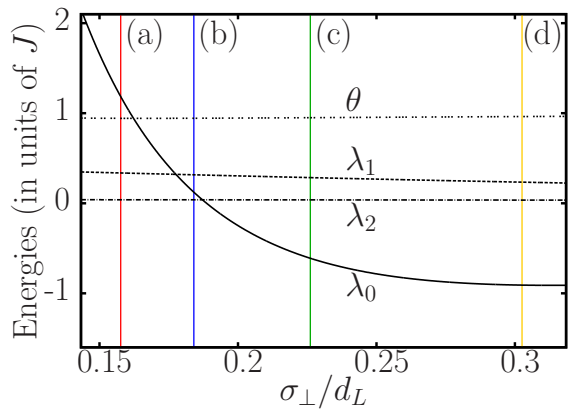}
\end{center}
\caption{\label{parameter} (Color online) Site energies
  $\lambda_0$, $\lambda_1$ and $\lambda_2$, and hopping energy
  $\theta$ as function of $\sigma_\perp$ in units of $d_L$, for the
  case of Dysprosium atoms with $D^2/d_L^3=0.016 E_R$, $\alpha=\pi/2$,
  $\lambda_L=2d_L=570$ nm, $s_{(T)}=6$ and $s_{(I)}=30$. The colored
  vertical lines labeled by different letters correspond respectively
  to: (a) $\sigma_{\perp}/d_L=0.158$, (b) $\sigma_{\perp}/d_L=0.184$,
  (c) $\sigma_{\perp}/d_L=0.226$, and (d) $\sigma_{\perp}/d_L=0.302$
  and identify the set of parameters that we used for the calculation
  of the localization properties of the system.}
\end{figure}

In Fig.~\ref{parameter}, we show the behavior of the quantities
$\theta=(J+J^d)/J$ and $\lambda_{n-l}=u^{dd}_{n-l}/J$ for $|n-l|=0,1$
and 2, for $D^2/d_L^3=0.016 E_R$, $\alpha=\pi/2$, $s_{(T)}=6$ and
$s_{(I)}=30$ as a function of $\sigma_{\perp}$ which is our control
parameter. This value of $D^2/d_L^3$ corresponds to the case of
Dysprosium atoms trapped in an optical lattice generated by a laser of
wavelength $\lambda_L=570$ nm \cite{Lev2011}.

We note that, for this choice of parameters, we can reasonably set
$\theta=1$ ($J^d=0$) and approximate the onsite energies for
$|n-l|\ge2$ by the asymptotic expression $\lambda_{n-l}=
\lambda/|n-l|^3$ with $\lambda=D^2/(Jd_L^3)$. They are therefore
independent on the value of $\sigma_\perp$.  Also $\lambda_1$ does not
depend significantly on the radial confinement. Conversely the value
of $\lambda_0$ strongly depends on $\sigma_\perp$ and can even vanish
and become negative due to the anisotropy of the dipolar interaction.

In the presence of several impurities the different contributions have
to be included summing over the occupied sites $\bar{l}$. Because of
the impurity-impurity dipolar repulsion, we will impose that each
impurity has to be preceded and followed by at least two empty
sites. The resulting Hamiltonian,
\begin{equation}
\begin{split}
H=&-J\sum_n  \left(\mid w_n\rangle \langle w_{n+1}\mid+ \mid w_{n+1}\rangle \langle w_n\mid\right)\\
&+ \sum_n \varepsilon_n \mid w_n\rangle\langle w_n\mid ,
\label{eq:tb_Hamiltonian_2}
\end{split}
\end{equation}
has hopping energies equal to $J$, and site energies
\begin{equation}
\varepsilon_n=\sum_{\bar{l}}u^{dd}_{n-\bar{l}}=\sum_l\rho_lu^{dd}_{n-l}.
\end{equation}

The system is characterized by the properties of the impurity density
distribution and the interaction potential. We introduce the notation
$\langle \dots \rangle$ to indicate the averaging for each lattice $l$
over different realizations of the disordered potential. Since, due to
translational invariance such averages will not depend on the lattice
site, the index $l$ will not appear in our notations for the
correlation functions.

The discretized impurity density distribution $\rho_l$ is a stochastic
variable with average value

\begin{eqnarray}
\langle\rho_l\rangle=\mathcal{C},
\end{eqnarray}
corresponding to the impurity concentration $\mathcal{C}$, and
density correlation function

\begin{equation}
C_\rho(\ell)=\langle\rho_l\rho_{l+\ell}\rangle.
\end{equation} 

The average value and the correlation functions of the full potential
$\varepsilon_l$ can be extracted from the statistical properties of
$\rho_l$ and the shape of the interaction potential. In particular,
one can prove two important relations. First that the average value of
the full potential is simply given by

\begin{equation}
\langle\varepsilon_l\rangle=\mathcal{C} \sum_nu^{dd}_{n},
\end{equation}
which is the product of the impurity concentration and $\sum_nu^{dd}_{n}$.
This last quantity can be thought as a sort of spatial average of the
interaction potential.
Second one can show that
\begin{equation}
C_\varepsilon(\ell)=\langle\varepsilon_l\varepsilon_{l+\ell}\rangle=\sum_jC_\rho(\ell-j)C_u(j),
\label{eq:convolution}
\end{equation}
namely that the two-point correlation function is given by the
convolution of the density correlation function $C_\rho(\ell)$ and

\begin{equation}
C_u(\ell)=\sum_n u^{dd}_n u^{dd}_{n+\ell}.
\label{eq:correlation_u}
\end{equation}
which is the interaction potential correlation function.

For the case of a dipolar potential $\lim_{\ell\rightarrow\infty}C_u(\ell)\propto\ell^{-3}$, and for random impurities whose minimum distance is fixed to be three sites, as in our model, one has

\begin{equation}
C_\rho(\ell)=\mathcal{C}^2+
\left(\frac{\mathcal{C}}{1-2\mathcal{C}}\right)^{\ell/2}
\left[A\cos(\kappa\ell)+B\sin(\kappa\ell)\right],
\end{equation} 
with $\kappa=d_L^{-1}[\pi-{\rm atan}\sqrt{(4-9\mathcal{C})/\mathcal{C}}]$, $A=\mathcal{C}-\mathcal{C}^2$ and $B=-[\sqrt{\mathcal{C}^3(1-2\mathcal{C})}+(\mathcal{C}-\mathcal{C}^2)\cos(\kappa d_L)]/\sin(\kappa d_L)$.

Thus we can conclude that the impurity distribution introduces short-range 
correlations, while the shape of the interaction $u_n^{dd}$ is responsible 
for long-range correlations. The role and the competition between these two 
effects will be extensively discussed in Sect.~\ref{sec:corr}. 

For the full potential, let us also introduce the reduced correlation function, defined as

\begin{equation}
	c_\varepsilon(\ell)=\frac{\langle\varepsilon_l\varepsilon_{l+\ell}\rangle - \langle\varepsilon_l\rangle^2}{  \langle\varepsilon_l^2\rangle - \langle\varepsilon_l\rangle^2  }
	\label{eq:reduced_correlation}
\end{equation}
and the associated spectral density

\begin{equation}
	S(k)=\sum_\ell c_\varepsilon(\ell) e^{ik\ell}.
	\label{eq:spectral_density}
\end{equation}
In the following, we will use the square root of the variance of the
full potential to quantify the potential strength
\begin{equation}
W=\sqrt{\langle\varepsilon_l^2 \rangle-\langle \varepsilon_l \rangle^2}.
\end{equation}

\section{Nature of the spectrum}
\label{thespectrum}

\begin{figure*}
\begin{center}
\includegraphics[width=0.8\textwidth]{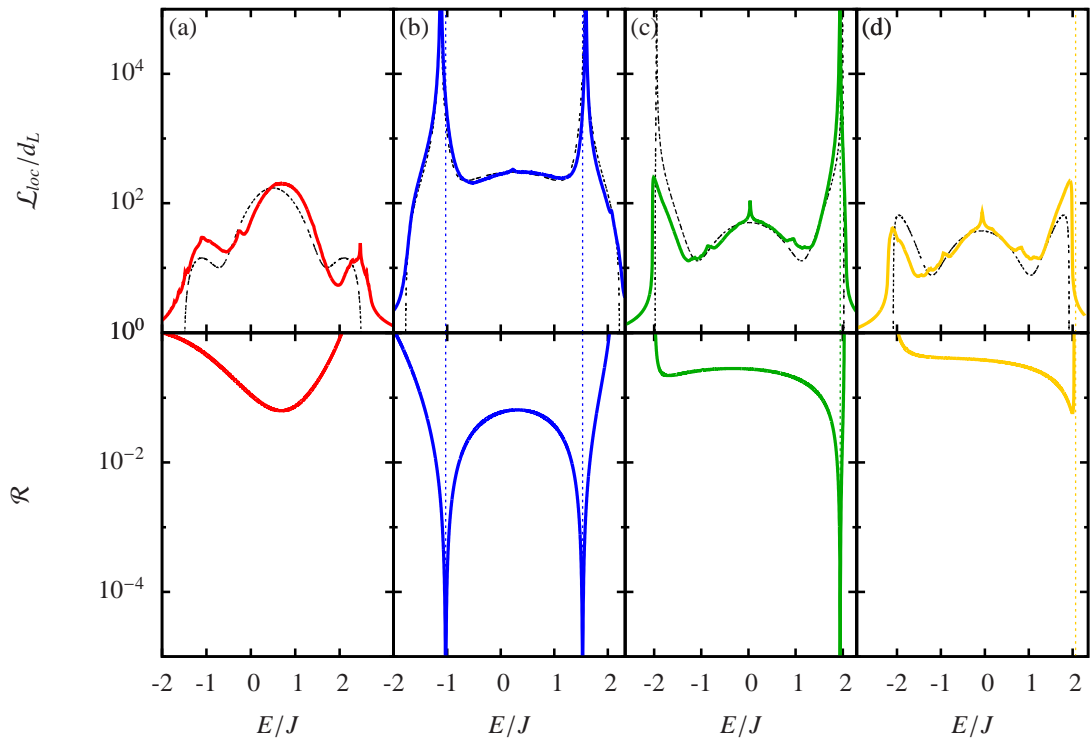}
\end{center}
\caption{\label{fig3} (Color online) Top panel: localization length
  $\mathcal{L}_{loc}$ in units of the lattice spacing $d_L$ as a
  function of the energy in units of $J$ for $\mathcal{C}=1/4$, system
  sizes up to $10^7$ lattice sites and averaged over $100$
  configurations.  The black dashed lines correspond to the
  localization length calculated in Born approximation
  $\mathcal{L}_{loc}^{(2)}$. Bottom panel: reflection coefficient
  $\mathcal{R}$ of the single impurity as a function of the energy in
  units of $J$.  The vertical dashed lines indicate the energies for
  which the reflection coefficient vanishes $\mathcal{R}(E)=0$. From
  left to right, different plots refer to increasing values of
  $\sigma_\perp$, corresponding to the vertical lines in
  Fig.~\ref{parameter}. }
\end{figure*}

We study the nature of the spectrum of the test-dipole by evaluating
the Lyapunov exponent $\Lambda(E)$ through the asymptotic relation 

\begin{equation}
\Lambda(E)=\frac{1}{\mathcal{L}_{loc}(E)}
=\lim_{N\rightarrow\infty}\dfrac{1}{Nd_L}\ln\left|
\dfrac{G_{N,N}(E)}{G_{1,N}(E)}\right|,
\label{eq-lyap}
\end{equation}
where $\mathcal{L}_{loc}(E)$ is the localization length,
${G(E)=(E-H)^{-1}}$ is the Green's function related to the Hamiltonian
$H$ at energy $E$, and $G_{i,j}(E)=\langle i|G(E)|j \rangle$.  The
matrix elements $G_{1,N}(E)$ and $G_{N,N}$, where $N$ is the total number of lattice sites, have been computed by
exploiting a renormalization-decimation scheme \cite{Farchioni1992a}.
Our results, obtained as the average over several configurations, are
shown in the upper row of Fig.~\ref{fig3}, where we consider
increasing values of $\sigma_\perp$, corresponding to the vertical
lines in Fig.~\ref{parameter}. We use the same color code in all
figures and label corresponding simulations with the same letters
$(a)$, $(b)$, $(c)$ and $(d)$.  Here and in the following, we consider
system sizes up to $10^7$ lattice sites, fix
$\mathcal{C}=1/4$, and average over $100$ 
configurations.  Each configuration is
generated by randomly distributing the impurities along the lattice and
forbidding those configurations where the minimum distance between
impurities is less than 3 lattice sites.  

As the value of $\sigma_\perp$ is increased, we observe very different
localization regimes. Notably for certain values of $\sigma_\perp$, we
observe divergences of the localization length, corresponding to the
appearance of metallic states in the spectrum. This suggests the
presence of delocalization effects induced by the correlations of the
physical model under consideration.
More precisely, for large positive values of $\lambda_0$, all states
are clearly localized since the localization length is always finite
(first panel, $(a)$). By increasing $\sigma_\perp$, for almost
vanishing values of $\lambda_0$ the localization length exhibits two
well defined peaks in two regions of the spectrum (second panel,
$(b)$). Increasing $\sigma_\perp$ further, corresponding to negative
values of $\lambda_0$, we observe the disappearance first of one of
the two divergences (third panel, $(c)$), and then of both of them
(fourth panel, $(d)$). In this last panel no divergences of the
localization length are observed, but there are still peaks at the
band edges, that recall the diverging behavior previously observed.

The dashed black lines correspond to the localization length
$\mathcal{L}_{loc}^{(2)}$ evaluated in Born approximation, which
corresponds to a second order perturbative calculation in the disorder
strength~\cite{Luck1989,Izrailev1999,Izrailev2012},

\begin{equation}
	\Lambda^{(2)}(E)=\frac{1}{\mathcal{L}_{loc}^{(2)}(E)}=\frac{W^2}{J^2} \frac{S\left(2 k(E) \right)}{ 8 \sin^2[k(E)d_L] }, 
	\label{born}
\end{equation}
where the connection between $k(E)$ and the energy is given by the
following relation $E=\langle\varepsilon_n\rangle+2J\cos(kd_L)$.  Let
us note that the Born approximation gives, by construction, a
symmetric localization length around the average value of the disorder
$\langle\varepsilon_n\rangle$, since the spectral density $S(k)$
associated to $c(\ell)$ is always a symmetric function of $k$. Despite
this fact, there is a noticeable agreement between the Born
approximation and the exact numerical results, even in case (c) of a
single divergence, where a strong peak at negative energy is strongly
reminiscent of the divergence found in the Born approximation.

\section{Role of correlations}
\label{sec:corr}

With the aim of understanding why we are observing the appearance and the
disappearance of metallic states in the spectrum by varying the radial 
confinement (and thus the effective 1D dipolar interaction), we analyse the effects
of the short-range correlations introduced by the impurity density
distribution and of the long-range correlations introduced by the dipolar
potential between the test dipole and the dipolar impurities. 

\subsection{Short-range correlations}

In order to isolate the role of short-range correlations in our model, we
calculate the reflection coefficient for the case of a single impurity, as in Eq.~(\ref{eq:tb_Hamiltonian}), trapped at the site of index $0$ of an infinite lattice. 
To get analytical results, we first neglect contributions beyond nearest neighbours. Therefore we assume that a single dipolar impurity modifies just a trimer of on-site energies $\{\lambda_1,\lambda_0,\lambda_1\}$ and we generally assume that it can modify also the hopping with nearest neighbouring sites $\theta$.

A plane wave $\langle n|k\rangle=e^{iknd_L}$, eigenstate of the unperturbed periodic Hamiltonian ${H^0=\sum_{n=-\infty}^{+\infty}}{  -J (\mid w_n\rangle \langle w_{n+1}\mid+\mid w_{n+1}\rangle \langle w_n\mid)}$ with energy $E=-2J\cos(kd_L)$, is perturbed by the impurity and results in the following wavefunction

\begin{equation}
\langle n|\varphi\rangle = \left\{
 \begin{array}{ll}
\tau e^{iknd_L} & (n>1)\\
e^{iknd_L}+r\, e^{-iknd_L} & (n<-1)
\end{array} \right.\,,
\end{equation}
where $\tau$ is the transmittance and $r$ the reflectance.
Using the scattering formalism combined to the renormalization-decimation scheme \cite{Farchioni1999a,Bakhtiari2005a,schaff10}, we obtain the following analytical formula for the reflection coefficient $\mathcal{R}=|r|^2$ of the single dipolar impurity

\begin{widetext}
\begin{equation}
\mathcal{R}=\dfrac{\left\{\lambda_1\left(\frac{E}{J}\right)^2-{\frac{E}{J}}[1-\theta^2+\lambda_1^2+\lambda_1\lambda_0]-2\theta^2\lambda_1+\lambda_0+\lambda_1^2\lambda_0\right\}^2}{\left\{\lambda_1\left(\frac{E}{J}\right)^2-{\frac{E}{J}}[1-\theta^2+\lambda_1^2+\lambda_1\lambda_0]-2\theta^2\lambda_1+\lambda_0+\lambda_1^2\lambda_0\right\}^2+\theta^4\left[4-\left(\frac{E}{J}\right)^2\right]}.
\label{eq:refl}
\end{equation}
\end{widetext}
In the bottom row of Fig.~\ref{fig3}, we plot $\mathcal{R}$ for the same parameters used for the calculation of the localization length, i.e. $\theta=1$ and $\lambda_0$ and $\lambda_1$ taken from the curves in Fig.~\ref{parameter}. 
We observe that the calculation of the reflection coefficient of the single impurity provides a very good understanding of the behavior of the localization length:  the energies where $\mathcal{R}$ tends to zero are exactly those where the localization length exhibits very large anomalous values. 
There is therefore a direct connection between the appearance of metallic states in the spectrum and the scattering properties of the single impurity.
It has been previously shown by Dunlap \emph{et al.}  \cite{Phil90} that this kind of single impurity analysis can be used to interpret the transport properties of a system of $N$ lattice sites, where there are several randomly placed impurities.
More precisely they proved that in such systems the number of single-particle states that show a metallic behavior, being extended over the full system, is of the order of $\sqrt{N}$. Notably this number of delocalized states is large enough to induce transport in the system and initially localized wavepackets show a superdiffusive spreading in the disordered potential. This means that this type of extended states are detectable in typical expansion experiments which can be performed with ultracold atomic gases \cite{Billy2008,Roati2008}. 

\begin{figure}
\includegraphics[width=0.85\linewidth]{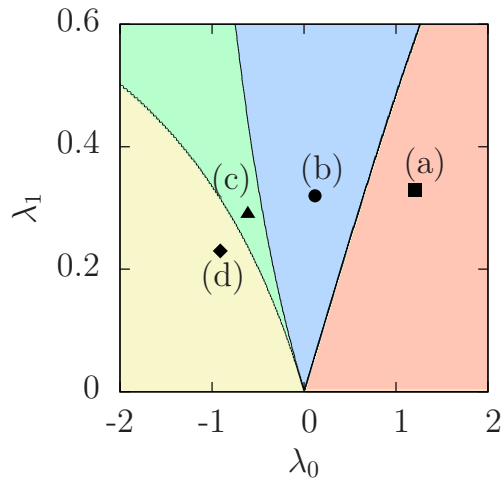}
\caption{\label{phase-dia} (Color online) Phase diagram induced by
  short-range correlations extracted from the reflection coefficient
  in Eq.~(\ref{eq:refl}) for the single impurity case. The different
  regions correspond to different localization regimes obtained from
  the solution of Eq.~\eqref{eq:zeros}. No real solutions of
  Eq.~\eqref{eq:zeros} correspond to the red region (a). If
  Eq.~\eqref{eq:zeros} has real solutions, we can distinguish three
  cases depending on the number of solutions lying inside the spectrum:
 two solutions (blue region (b)), one solution (green region (c)), no solutions (yellow region,
  (d)). The four markers in the diagram correspond to the simulations
  presented in Fig.~\ref{fig3} and to the values of $\sigma_\perp$
  indicated by vertical lines in Fig.~\ref{parameter}.}
\end{figure}

It is remarkable that making use of the simple analytical expression \eqref{eq:refl} we can predict the localization properties of a rather complex system and the occurrence of metallic states in the spectrum. Studying the solutions of the equation 
\begin{equation}
	\mathcal{R}(E)=0 
	\label{eq:zeros}
\end{equation}
as a function of $\lambda_0$ and $\lambda_1$, one can extract the phase diagram in Fig.~\ref{phase-dia}.
We identify four different regions depending on the
number of solutions of Eq.~\eqref{eq:zeros} and on their values.  More
precisely, if the solutions are both imaginary, no divergences are
expected and all the states are exponentially localized (red region
(a)). If the solutions are real and inside the single impurity
spectrum $E=-2J\cos(k d_L)$, namely when the roots satisfy the
additional condition $|E|<2J$, divergences are expected.  Therefore,
when the solutions are real, we can identify three additional
scenarios: both solutions lie inside the spectrum (blue region (b)),
only one solution lies inside the spectrum (green region (c)), both
solutions lie outside the spectrum (yellow region (d)).

In the diagram shown in Fig.~\ref{phase-dia}, we identify with markers
the values of $\lambda_0$ and $\lambda_1$ corresponding to the
different plots of Fig.~\ref{fig3}, where the dashed vertical lines
mark the energies which verify the condition $\mathcal{R}(E)=0$.  In
particular the square in the red region $(a)$ corresponds to the first
plot in Fig.~\ref{fig3} where all states are localized; the circle in
the blue region $(b)$ corresponds to the second plot in
Fig.~\ref{fig3}, where we observe two resonances; the triangle in the
green region $(c)$ corresponds to the third plot in Fig.~\ref{fig3},
where we observe one resonance; and the diamond in the yellow region
$(d)$ corresponds to the last plot in Fig.~\ref{fig3}, where there are
no resonances but the peak on the right shows a tendency to diverge
due to the fact that the resonance lies just outside the single
impurity spectrum.

Let us recall that, in our single impurity analysis, we considered the
case where one isolated dipole induces just a trimer of site energies
$\{\lambda_1,\lambda_0,\lambda_1\}$ and we neglected beyond nearest
neighbour contributions. In other words, we neglected the effect of
long-range correlations.

In the next section, we study in detail the role played by the dipolar
tails that we neglected in this simplified calculation and we
highlight the role played by long-range correlations.

\subsection{Long-range correlations}

In order to understand the role played by long-range correlations and
place the dipolar case in a wider context, we investigate the
localization properties of a disordered potential generated by an
effective impurity-test particle interaction with tails decaying as
$u^\beta_{|n|}\sim1/|n|^\beta$ where $\beta\ge1$.

This is done by placing the impurities exactly as done in the dipolar
case, keeping fixed the values of $\lambda_0$ and $\lambda_1$ and
choosing $\lambda_n=u^\beta_n/J=\lambda/|n|^\beta$ for $n\ge2$.  The
case $\beta=3$ recovers our physical model with dipolar interactions.
Moreover, we shift and normalize the on-site energies in order to
obtain the same average value $\langle\varepsilon_n\rangle$ and
disorder strength $W$ that we had in the dipolar case. Following this
procedure, we can really analyze the effect of long-range correlations
keeping fixed the disorder strength $W$.  In particular we considered
values of $\beta$ ranging from $1$ up to $5$ and we also considered
the case of $\beta=\infty$ that corresponds to $\lambda_n=0$ for
$|n|\ge2$.

The potential generated with this procedure has $C_\rho(\ell)$ which
is unchanged and decays exponentially as previously discussed. This is
due to the fact that the impurities are placed exactly in the same way
as before. The correlation function associated to the interaction
potential $C_u(\ell)$ is instead modified and using
Eq.~\eqref{eq:correlation_u} one can show that it decays at large
distances as $C_u(\ell)\sim 1/\ell^\beta$ for $\beta>1$ and as
$C_u(\ell)\sim \ln(\ell)/\ell$ for $\beta=1$. These asymptotic
expressions determine the shape of the tails of the two-point correlation function
$C_\varepsilon(\ell)$ of the random potential seen by the test dipole
(see Eq. (\ref{eq:convolution})) and consequently the
reduced correlation function associated to the full potential
$c_\varepsilon(\ell)$.

\begin{figure}
\includegraphics[width=0.99\linewidth]{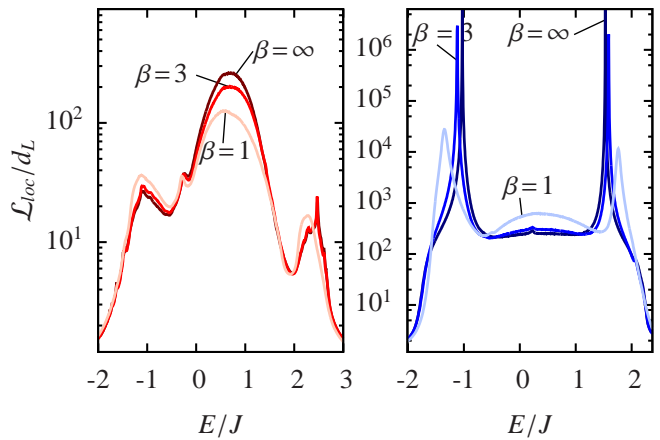}
\caption{\label{fig:long_corr} (Color online) Localization length
  $\mathcal{L}_{loc}$ in units of the lattice spacing $d_L$ as a
  function of the energy in units of $J$ for $\beta=1,3,\infty$,
  namely different types of long-range correlations identified by the
  asymptotic decay of the tails of the two-point correlation function
  $C_{\varepsilon}(\ell)$. The left and right panels correspond to two different
  localization regimes induced by short-range correlations as shown
  Fig.~\ref{fig3}(a,b) for $\beta=3$ (see text for more details).}
\end{figure}
The effects played by long-range correlations are again studied by
calculating numerically the localization length with the
renormalization-decimation approach. In Fig.~\ref{fig:long_corr} we
show the localization length $\mathcal{L}_{loc}$ calculated for
different values of $\beta$. In particular we show a comparison
between the two limiting cases of $\beta=\infty$ and $\beta=1$ and the
physical case under consideration, i.e. the dipolar case $\beta=3$.
We considered also other values of $\beta$ but we do not show the
results here since they are not particularly instructive. They just
show an intermediate behavior between the two limiting cases reported
here.

The two panels in Fig.~\ref{fig:long_corr} correspond to the two
different set of parameters already used for Fig.~\ref{fig3}(a,b). In
the left panel we show the case where the localization length is
always finite, while in the right panel we show the case where there
are two resonances in the spectrum. Therefore the two curves for
$\beta=3$ are exactly the same curves which are shown in the upper row
of Fig.~\ref{fig3}(a,b). We compare them with the case of complete
absence of long-range correlations ($\beta=\infty$) and with the case
of very slowly decaying correlations ($\beta=1$).

In Fig.~\ref{fig:long_corr}, there are two main features that we would
like to stress here. The first is the effect on the divergencies in
the localization length, discussed in the previous section. We observe
that such divergencies tend to be beveled by long-range correlations.
In fact for $\beta=\infty$, the localization length takes values of
the order of the system size ($10^7 d_L$) signaling the presence of
real metallic states in the system which extend over the full
lattice. As the value of $\beta$ is reduced, corresponding to slower
decaying tails, peaks in the localization length are still present but
they are shifted towards the band edges and their height is
decreased. This behavior is somehow expected, since the perfect
resonance condition, obtained with the single impurity calculation
presented in the previous section, is no more fulfilled in the
presence of several impurities with overlapping slowly decaying
tails. The tails tend to restore destructive interference in the
forward direction and thus introduce localization in the
system. However, from the results shown in Fig.~\ref{fig:long_corr},
we can conclude that the effect of short-range correlations remains
clearly visible also in presence of long-range correlations.

The second feature that we would like to highlight is the
counterintuitive behavior of $\mathcal{L}_{loc}$ introduced by
long-range correlations at the center of the band. In fact, depending
on the set of parameters under consideration, long-range correlations
have opposite effects on the localization length, which can either
decrease (left panel), or be enhanced (right panel). This observation
shows the highly nontrivial role played by long-range correlations in
determining the localization properties of a disordered system and
indicates a richer behavior with respect to what has been observed so
far in the literature \cite{croy2011}.

These features that we extracted from the numerical simulations
reported in Fig.~\ref{fig:long_corr} are also captured within the Born
approximation. We do not report the curves for
$\mathcal{L}_{loc}^{(2)}$ calculated in Born approximation, and just
comment that the agreement between those curves and the exact
numerical results is good, similar to that observed in
Fig.~\ref{fig3}.

One can understand the previous results based on the following
properties: (i) in the Born approximation, $\Lambda^{(2)}$ is
proportional to $S(2k)$ and to $W^2$ which we have taken to be
constant (see Eq. \ref{born}); (ii) the integral of $S(2k)$ is a
constant; (iii) the longer the range of the correlations, the larger
$S(2k)$ will be at the energy band edges.  In the case of
Fig.~\ref{fig:long_corr}~(right), where for $\beta=\infty$ there are
extended states, one can conjecture that, for decreasing $\beta$,
property (iii) above together with the desappearence of the extended
states lead to a decrease of $\Lambda^{(2)}$ at the energy band
center, corresponding to an increase of the localisation length, as
observed.  The case shown in Fig.~\ref{fig:long_corr}~(left) cannot be
explained based on similar simple arguments, since the increase of
$S(2k)$ at the energy band edges implies a non trivial redistribution
of the disorder spectral components all over the band. In the specific
case, one observes an increase of $\mathcal{L}_{loc}$ at the energy
band center, contrary to previous predictions \cite{croy2011}.

Finally we would like to comment that we do not find the presence of
mobility edges induced by long-range correlations as suggested in
\cite{deMour98,Garcia2009}.

\section{Conclusions}
\label{concl}
In this paper, we considered a set of dipolar impurities pinned at
random positions in a deep optical lattice which create a disordered
potential for an atom in a different internal state. An analysis of
the statistical properties of the model showed that repulsive dipolar
interactions between impurities introduce short-range correlations due
to the fact that occupations of neighboring sites are forbidden. 

The localization properties of the model were calculated by means of a
renormalization-decimation technique which allowed us to calculate
properties of very large systems and study the extended or localized
nature of the states. We found that the presence of short-range
correlations can give rise to different regimes. In particular, as the
parameters of the system are changed, we observed regimes where one or
more discrete sets of extended states appear in the spectrum. The
occurrence of the different regimes can be predicted starting from an
analytical expression obtained from the scattering of a single
impurity. 

Long-range correlations were studied not only for the dipolar case but
also for a more general two-point correlation function decaying as
$C(\ell)\sim 1/\ell^{\beta}$, where the case $\beta=3$ corresponds to
the dipolar case. We saw that long-range correlations in general tend
to restore localization in the spectrum, but also lead to
counterintuitive behaviors of the localization length. More precisely,
depending on the regime under consideration, they can enhance or
reduce localization at the center of the band.  

Our work sheds light on the interplay between the role of short-range
and long-range correlations and can be a guide for experiments devoted
to the study of Anderson localization with ultracold dipolar
gases. Natural extensions of the present work include the study of 2D
geometries and the role of interactions between many test dipoles.

\begin{acknowledgments}
This work was supported by Grants No. CNRS-24543 and TUBITAK-210T050,
by the LIA FSQL, by ERC through the QGBE grant and by Provincia
Autonoma di Trento.  We are grateful to Franco Dalfovo, Iacopo
Carusotto, Jean-Fran\c{c}ois Schaff and Luca Tessieri for useful
discussions.
\end{acknowledgments}


\end{document}